\documentclass[final,1p,times]{elsarticle}

\usepackage[utf8]{inputenc}
\usepackage{amsmath}
\usepackage{subcaption}
\usepackage{graphicx}
\usepackage{lineno}
\usepackage{url}
\usepackage{color}
\usepackage{gensymb}

\usepackage [english]{babel}
\usepackage [autostyle, english = american]{csquotes}
\MakeOuterQuote{"}

\usepackage{amssymb}
\usepackage{amsthm}
\usepackage{lineno}

\journal{NIM-A}
\title{Electrode Design for a Cavallo High Voltage Multiplier in a Cryogenic nEDM Experiment}

  \author[add1,add2]{Marie A. Blatnik}
  \ead{mblatnik@lanl.gov}
  \author[add2]{Steven M. Clayton}
  \ead{sclayton@lanl.gov}
  \author[add1]{Bradley W. Filippone}
  \ead{bradf@caltech.edu}
  
  \author[add2]{Takeyasu M. Ito}
  \ead{ito@lanl.gov}

  \author[add2]{Nguyen S. Phan}
  \ead{nphan@lanl.gov}
  
  \author[add2]{Christopher M. O'Shaughnessy}
  \ead{chriso@lanl.gov}

  \author[add3]{John C. Ramsey}
  \ead{ramseyjc@ornl.gov}

  \address[add1]{W. K. Kellogg Radiation Laboratory, California Institute of Technology,
Pasadena, California 91125, U.S.A.}
  \address[add2]{Los Alamos National Laboratory, Los Alamos, New Mexico 87545, U.S.A.}
  \address[add3]{Oak Ridge National Laboratory, Oak Ridge, Tennessee 37830, U.S.A.}

\date{April 7, 2026}

\begin{document}

\begin{frontmatter}           

\begin{abstract}
The Cavallo multiplier\cite{cavallo_complete_1795} is an electrostatic inductance machine that can generate low-noise high voltages electrically isolated from its voltage input, making it ideally suited for precision experiments. Its in-situ production makes it especially useful in cryogenic experiments, where the use of traditional feedthroughs is challenging due to thermal, electrical, magnetic, and physical size considerations. One such experiment is a cryogenic measurement of the neutron electric dipole moment (nEDM)\cite{ahmed_new_2019,itoHV_now}, which requires several hundred kilovolts on a  measurement cell electrode in 0.4 K liquid helium (LHe). A Cavallo multiplier can generate this voltage by stepping up a smaller input (e.g., 50 kV) from a feedthrough. We designed Cavallo electrodes using finite element analysis to provide high voltage gain and low probability of electrical breakdown in the experimental apparatus. The final geometry achieves a gain of 18, providing a target of 650 kV with peak electric fields of 116 kV/cm distributed over small areas to limit breakdown risk.
\end{abstract}


\end{frontmatter}


\section{Introduction}
The Cavallo multiplier is an electrostatic induction machine that achieves voltage multiplication through repeated charge transfer using a moving electrode. It multiplies an applied voltage without electrical contact between its input and output, allowing high voltage production with extremely low electrical noise on the isolated output. Its output is free from the electromagnetic interference and voltage ripples that typically accompany conventional high voltage sources. This makes the Cavallo multiplier especially useful in precision measurements where even small electrical disturbances can obscure sensitive signals. A further advantage of the Cavallo multiplier is its capability for in-situ high voltage generation. This feature is especially useful in cryogenic and ultra-low-noise experiments, where the implementation of traditional high-voltage feedthroughs presents significant challenges. 

An example of an experiment with these constraints is the cryogenic neutron electric dipole moment measurement \cite{ahmed_new_2019}, which requires hundreds of kilovolts in a magnetically precise, sub-kelvin region in liquid helium (LHe). Feeding this high voltage directly into the cryostat would cause unacceptable heat load on the sub-kelvin region by the leakage currents on the surfaces of the feedthrough insulator, and also by heat conduction through the electrical conductor from outside the cryostat. Identifying suitable nonmagetic materials also pose design challenges. The Cavallo multiplier was proposed to provide the required voltage to the high voltage electrodes for this nEDM measurement \cite{clayton_cavallos_2018}. 

\begin{figure}
    \centering
    \begin{subfigure}{0.3\textwidth}
        \label{ref_label1}
        \includegraphics[width=\linewidth]{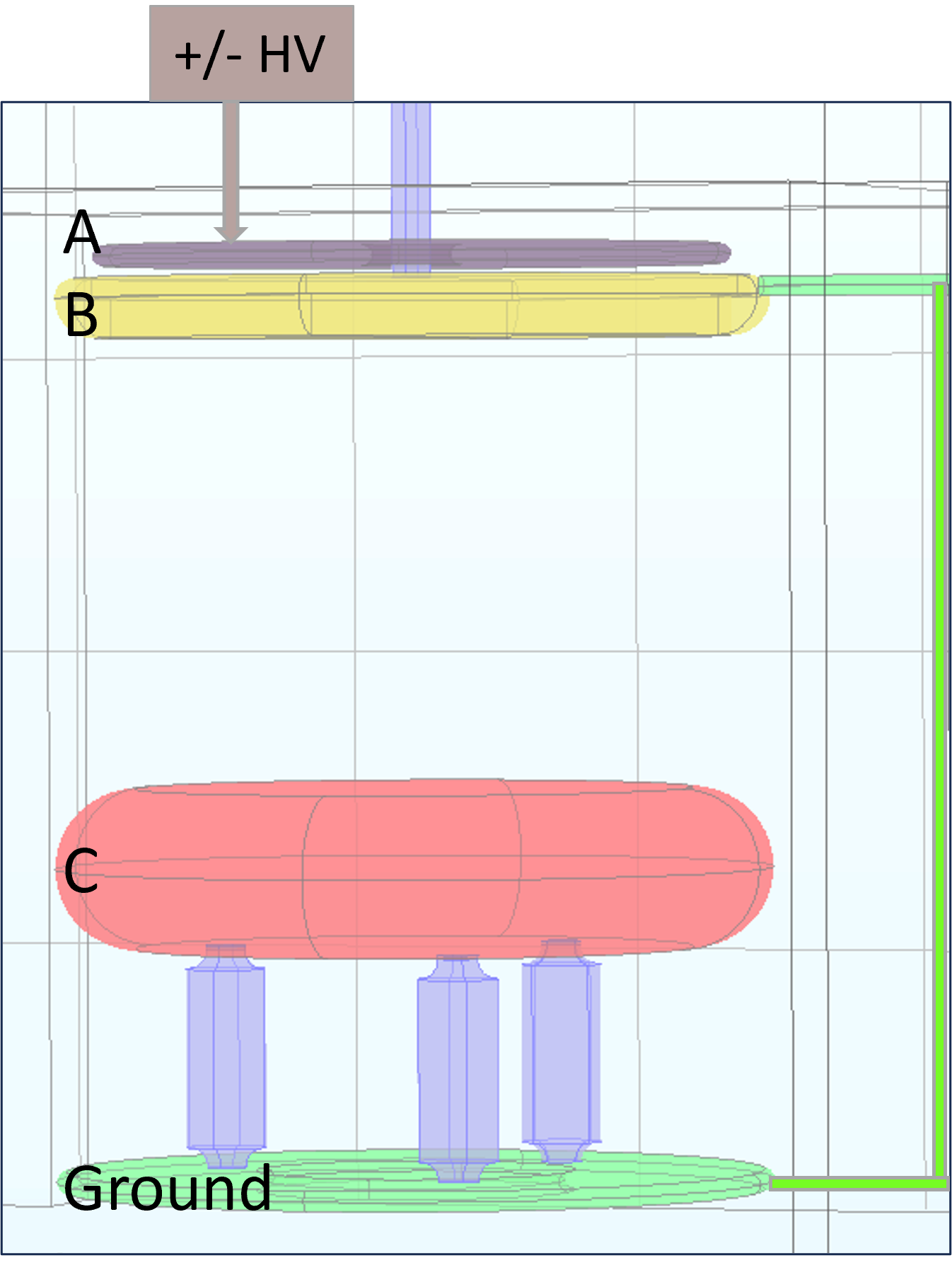}
        \caption{Position 0: Induce charge on B}
    \end{subfigure}
    \hspace*{\fill}
    \begin{subfigure}{0.3\textwidth}
        \label{ref_label2}
        \includegraphics[width=\linewidth]{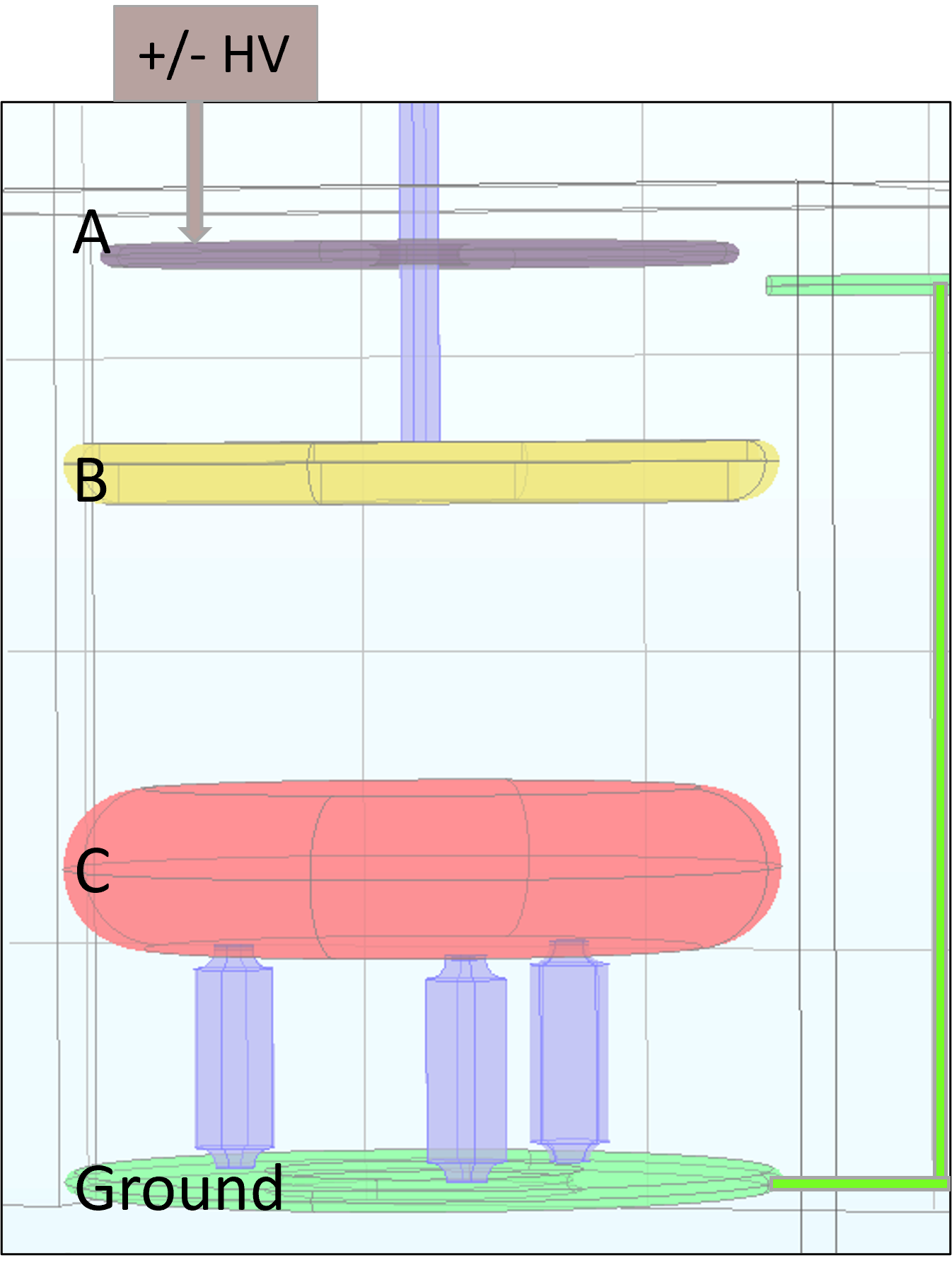}
        \caption{Move charge}
    \end{subfigure}
    \hspace*{\fill}
    \begin{subfigure}{0.3\textwidth}
        \label{ref_label3}
        \includegraphics[width=\linewidth]{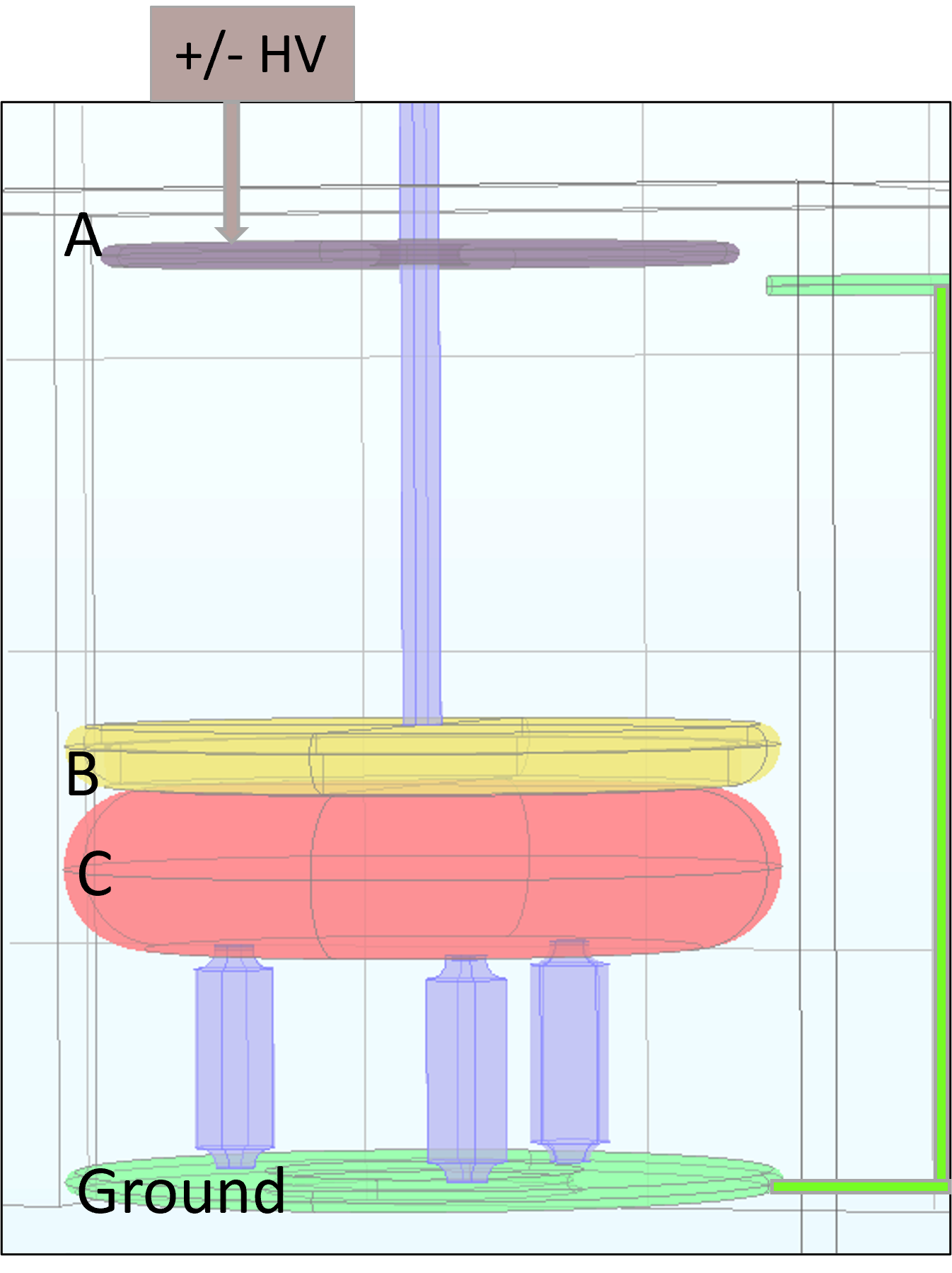}
        \caption{Position 1: Transfer charge to C}
    \end{subfigure}
    
    \caption{An example of a Cavallo multiplier: The charge is transferred to the B electrode in (a), then physically moved with the B electrode in (b), and transferred to the C electrode in (c). Then the B electrode returns to the position in (a) to recharge and repeat the cycle.}
    \label{fig:Cavallo_Conceptual}
\end{figure}

The Cavallo multiplier consists of three electrodes with an additional ground electrode, which are used to multiply a small input voltage using capacitive induction. A conceptual model is shown in Fig. \ref{fig:Cavallo_Conceptual}. The top electrode (A) is connected to an external power supply\footnote{A smaller 50 kV feedthrough compatible with the cryogenic requirements for the nEDM experiment is commercially available \cite{phan_study_2021,ito_apparatus_2016}, and a version of this feedthrough satisfying the nonmagnetic requirements is straightforward without significant research and development effort.} that provides a modest input voltage. The third electrode (C) is the high voltage electrode---the electrode in our application that must be charged to 650 kV. It electrically connects to the high voltage electrode of the measurement cells, which creates the near-uniform electric field inside the cells. 

The second electrode (B) moves between two positions to shuttle charge onto electrode C. In one position, depicted in Fig. \ref{fig:Cavallo_Conceptual}(a), electrode B is placed close to electrode A, where it electrically contacts a grounded conductor. In this position, electrode B gains an induced charge due to the capacitance formed by electrodes A and B. Electrode B then disconnects from ground, retaining its charge, and moves toward electrode C (depicted in Fig. \ref{fig:Cavallo_Conceptual}(b)). It touches electrode C in the second position (depicted in Fig. \ref{fig:Cavallo_Conceptual}(c)), where the charge redistributes in proportion to the capacitances that electrodes B and C form with the rest of the system. With no initial charge on the C electrode, most of the charge moves from electrode B to electrode C due to the large capacitance electrode C forms with the ground. Electrode B then returns to its starting position (grounded near electrode A), completing its cycle and starting the next one. 

The final theoretical gain of the Cavallo multiplier as a function of the electrode capacitances is given by the following expression \cite{clayton_cavallos_2018}:
\begin{equation}
    G_C^1 = \frac{C_{AB}^1 - C_{AB}^0}{C_{AB}^1+C_{BG}^1+C_{BC}^0 C_{CG}^1/(C_{BC}^0+C_{CG}^0)}
    \label{eq:gain}
\end{equation}
where $C_{ij}$ denotes the mutual capacitance between the \textit{i}-th and \textit{j}-th electrodes, and the small capacitance $C_{AC}$ is neglected. The superscripts denote whether the capacitance is measured in "position 0," with electrode B grounded near electrode A as in Fig. \ref{fig:Cavallo_Conceptual}(a) or in "position 1" with electrodes B and C touching as in Fig. \ref{fig:Cavallo_Conceptual}(c). This gain is negative; the voltage on electrode C builds up with the opposite charge to that on electrode A.

This apparatus is limited both by the capacitances in Eq. (\ref{eq:gain}) (the "theoretical gain") and by the ability to withstand the electric fields in operation. Note how the B electrode capacitances ($C_{BG}^1+C_{AB}^1$) limit the fraction of charge transferred to the C electrode as they touch, limiting the gain. The $C^0_{BC}$ also limits the gain as the voltage increases on the C electrode by reducing the charge loaded onto the B electrode. However, the gain in Eq. (\ref{eq:gain}) neglects the potential for electrical breakdown in the insulating material surrounding the electrodes, which can limit the gain before Eq. (\ref{eq:gain}) does.

\begin{figure}
    \centering
    \includegraphics[width=\linewidth]{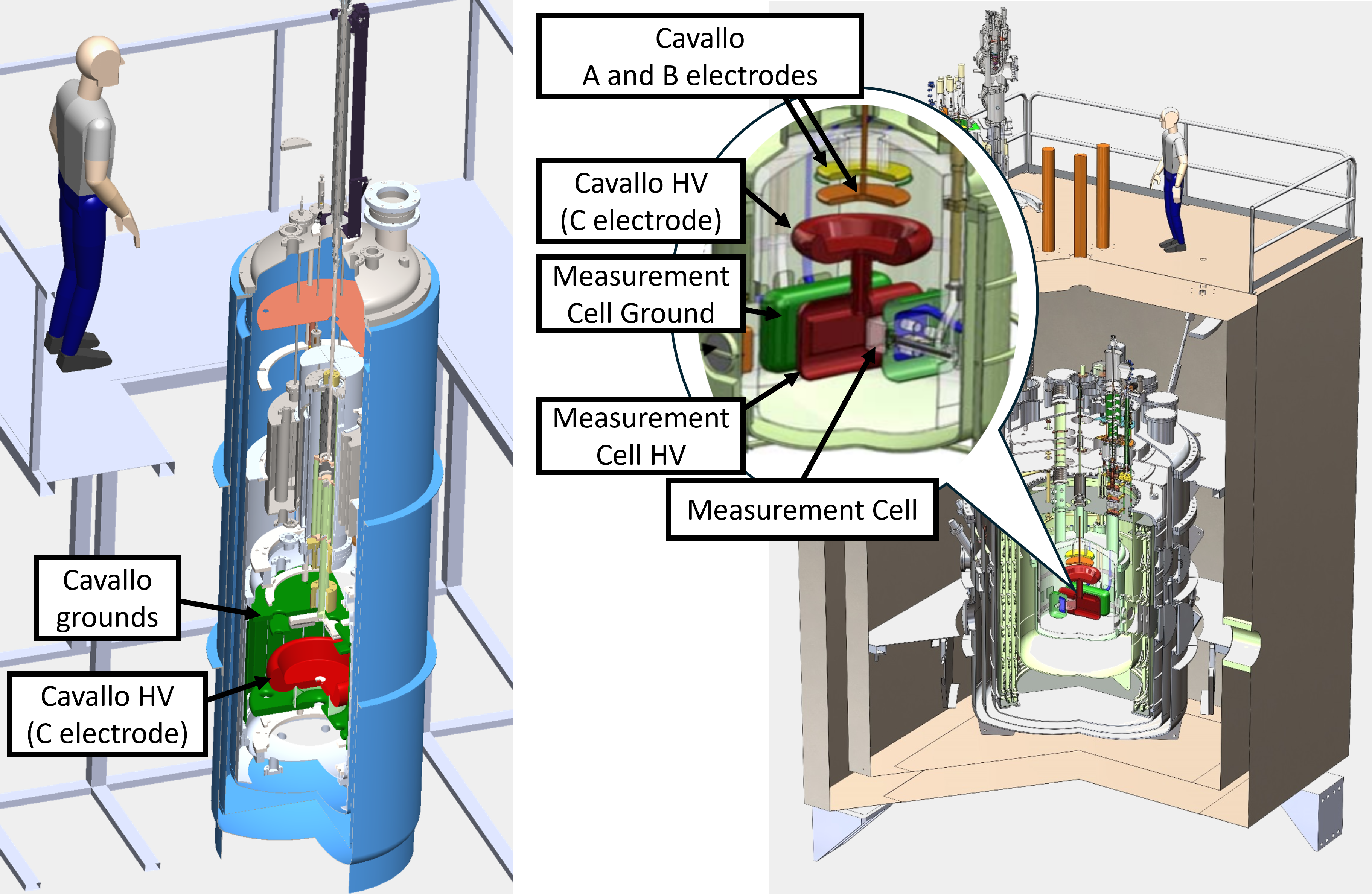}
    \caption{Computer-aided drawings showing the Cavallo multiplier in situ. The left panel depicts the apparatus installed within the test cryostat (the focus of this paper), with the Cavallo multiplier located in the lowest inner volume. The right panel shows the Cavallo multiplier in the heart of the nEDM experiment, with the Cavallo multiplier and the attached nEDM cell electrodes enlarged in the bubble. The C electrode is connected to the high voltage electrode of the nEDM measurement cells.}
    \label{fig:Cavallo_Setup}
\end{figure}

Reference~\cite{clayton_cavallos_2018} demonstrated that nEDM measurement electrodes can be paired with a Cavallo multiplier design capable of reaching 700 kV in 20 cycles while maintaining a sufficiently low maximum surface field. Based on this proof of principle, the goal of the present study was to develop a set of construction-ready electrodes for the Cavallo apparatus with high voltage reach while maintaining acceptably low breakdown risk. Our design process began with trial shapes described by parameterized curves. The shapes were then iteratively refined to increase the gain, while keeping peak surface electric fields below 120 kV/cm. This threshold was adopted as a simple design criterion based on our earlier study \cite{ito_apparatus_2016}. 

This design was tailored to the smaller test apparatus shown in the left panel of Fig.~\ref{fig:Cavallo_Setup} while preserving physical constraints relevant to the nEDM experiment. In this configuration (left panel), the Cavallo multiplier is located within the inner cryogenic volume at the base of the apparatus, with the C electrode shown in red. The green electrodes are the grounds in the apparatus, with the distance to the bottom electrode setting the load capacitance approximately equal to that of the nEDM measurement cell electrodes. These full-scale electrodes can be placed into the cryogenic nEDM experiment, shown on the right. The inset shows an enlarged view of the electrodes, highlighting the connection between the C electrode and the measurement cell electrodes.

Finally, we then confirmed that our design had an acceptably small electrical breakdown probability, specifically that the breakdown probability for the Cavallo electrodes was less than the breakdown probability of the nEDM measurement cell electrodes. Phan et al. \cite{phan_study_2021} showed that electrical breakdown depends not only on the field strength at the surface of an electrode, but also on the integrated surface area at that field strength. We applied their method, which finds the breakdown probability for arbitrary electrode geometries using breakdown field data from reference electrodes that include the other parameters that affect electrical breakdown (material, surface quality, LHe pressure, etc.), to our design. The analysis used available data on electropolished stainless steel electrodes at different LHe pressures. This material was selected for the initial full-scale prototype electrodes due to its durability, cryogenic properties, and established performance in similar applications \cite{ito_apparatus_2016,phan_study_2021}. These calculations are expected to apply to electrodes made of conductive-coated plastic and bulk resistive material, candidate materials for the future experimental apparatus \cite{itoHV_now}.


\section{\label{sec:methods}Methods}

\begin{figure}
    \centering
    \includegraphics[width=0.9\linewidth]{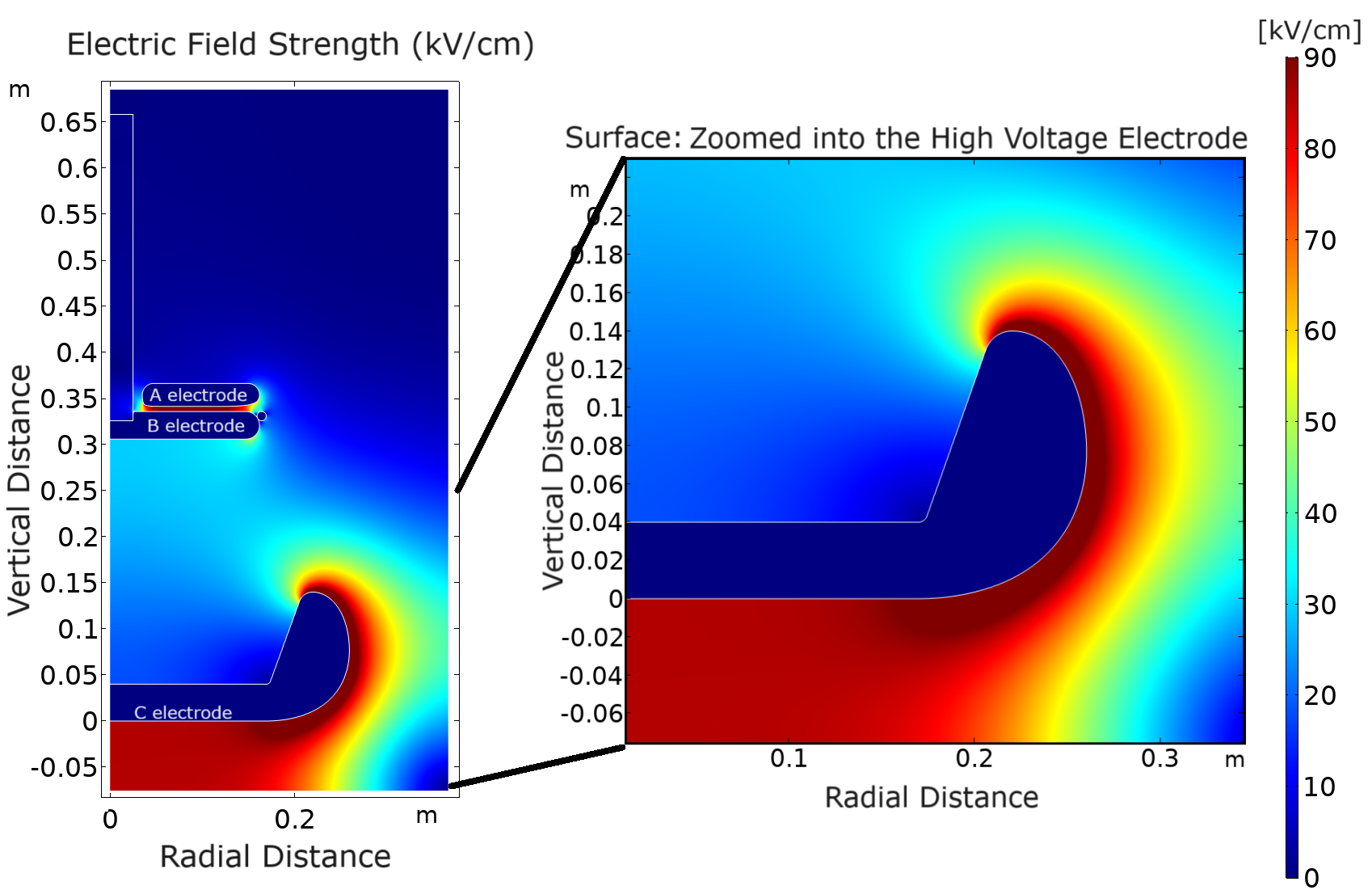}
    \caption{The axisymmetric electrostatic model for the Cavallo electrodes. The electrode shapes were tailored together as one system. Left: the C electrode at 650 kV within the electrostatic model. Right: the C electrode in detail---created with a large lobe as a result of the relationship between $C_C/C_B$ and $Q_C/Q_B$, maximizing the theoretical gain and reducing the electric field maxima region with two smoothly joined parametric curves.
    }
    \label{fig:1}
\end{figure}

The design goal for the Cavallo multiplier electrodes in the test apparatus is to achieve a sufficiently high gain ($\gtrsim$13) to reach the necessary 650 kV from 50 kV in less than 20 cycles, while reducing the probability of breakdown. The constraints include:
(1) volume constraints imposed by cryogenic nEDM experimental engineering considerations, and (2) a simple threshold of 120 kV/cm as discussed previously to keep the electrode design process efficient, with subsequent verification using the breakdown probability method of Phan et. al to ensure a sufficiently high limiting voltage. 

These goals were pursued using the finite element simulation software COMSOL \cite{noauthor_comsol_2019} to generate candidate electrode profiles for the test apparatus. The design process began from a baseline electrode geometry motivated by electrostatic considerations, which was then refined through iterative adjustment of the electrode curvature as described below. For each iteration, we calculated the capacitances, voltage distributions, and electric field maps of the apparatus. The refinement process focused on reducing regions of high electric field stress by dispersing electric field lines, which lowered the predicted probability of electrical breakdown.

 Two kinds of COMSOL simulations were run: 
 \begin{enumerate}
     \item Two-dimensional (2D) axisymmetric simulations: Models computed in 2D that represent three-dimensional space by rotating around the vertical (z) axis as in Fig. \ref{fig:1}. These models informed the electrode shapes and engineering constraints, and working in two dimensions is easier to visualize and saves computation time.

    \item Full three-dimensional (3D) simulation: the axisymmetric geometry results informed a Solidworks \cite{noauthor_dassault_nodate} engineering model imported into COMSOL, whose symmetry was broken by only a few features (e.g., mounting hardware on the B electrode, and the grounded housing cylinder for the apparatus was broken into discrete facets). This model better represented the future test apparatus in physical reality and verified that the engineering changes did not compromise the electrostatic design. 
     
 \end{enumerate}

Design considerations for each electrode were driven by its specific purpose, contribution to the gain, and areas of high electric field stress. The A electrode needed a central hole to accommodate the driving shaft of the B electrode -- the cylindrical rod that connects the B electrode to an external actuator. Higher gain is achieved by reducing the distance between the A electrode and the initial position of the B electrode (increasing $C_{AB}^0$), which creates a region of high electric field between them, especially at the edges; the bottom surface of the A electrode (facing the B electrode) needed sufficient refinement. 

\begin{figure}[!t]
  \begin{subfigure}{0.5\textwidth}
    \includegraphics[width=\linewidth]{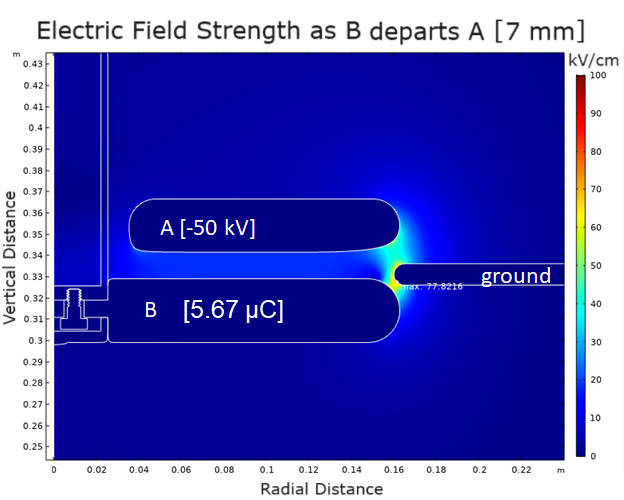}
    \caption{B electrode approaches the ground ring} 
  \end{subfigure}%
  \hspace*{\fill}   
  \begin{subfigure}{0.5\textwidth}
    \includegraphics[width=\linewidth]{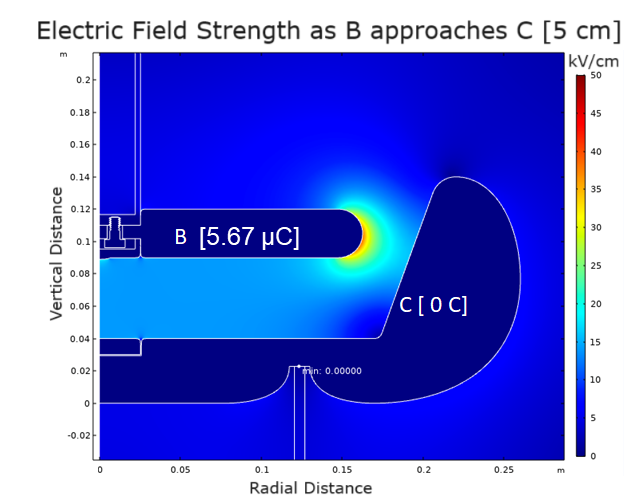}
    \caption{B electrode approaches the C electrode} 
  \end{subfigure} 
\caption{Pseudo-axisymmetric simulations of the electric field strength for two positions of the B electrode. (Pseudo-axisymmetric implies that these screws are suggestive only, and are not correct in the
2D-axisymmetric framework.)
The B electrode experiences high electric field on both its upper and lower faces as it traverses the distance between the ground ring and the C electrode; its best edge shape is a circular fillet.} \label{fig:Bapproaches}
\end{figure}

Electrode B was designed with rounded edges (constant-radius edge curvature) to mitigate the dynamic electric-field stress it experiences during operation. Its top surface sees a high electric field when it approaches the ground ring near A as in Fig. \ref{fig:Bapproaches}(a), and its bottom surface experiences higher electric fields when it approaches the C electrode as in \ref{fig:Bapproaches}(b). To maximize the radius of curvature of the electrode cross section at both the top and bottom edges, the solution is a circular fillet.






Further considerations for the B electrode include preventing electrical breakdown at either end of its travel. When the B electrode returns to the ground ring, its residual charge creates a potential difference between them, even if the A electrode is set to zero. Breakdown arising from this potential difference can be suppressed operationally by either adjusting the A electrode voltage or biasing the ground ring. In the former approach, the A electrode voltage is adjusted to eliminate the potential difference; the residual charge on the B electrode equals the charge required by the capacitance between the A and B electrodes. In the latter option, the ground ring--being isolated from other grounded surfaces--can be biased by an independent power supply to match the residual voltage of the B electrode until physical contact. Restoring the A or ground voltages after contact is established ensures that charge transfer occurs through conduction rather than electrical discharge.


Preventing a discharge when the B electrode approaches the C electrode is only theoretically possible. The C electrode must be flat where the B electrode makes contact. A flawless, perfectly parallel set of infinite planes will not produce any electrical discharge (a spark) in the medium \cite{clayton_cavallos_2018}. In reality, a spark between these two surfaces is inevitable due to imperfections. We have elected to localize the discharge to a replaceable button threaded into the bottom of electrode B and the top of electrode C (see Sec. \ref{section:BCbreakdown}).

The basic shape of the C electrode combines a lobe with an additional feature to increase the gain by screening the B electrode from the ground when it touches the C electrode. As the ratio $C^1_{CG}/C^1_{BG}$ increases, the ratio $Q_C/Q_B$ increases correspondingly. Sufficient geometric refinement is required to minimize the electric field at the outer edge of the lobe and support operation at 650 kV. 

We introduced an additional (fourth) electrode---designated the D electrode---by isolating the ground return beneath the C electrode, providing an option to measure leakage currents from the C electrode. The distance between the C electrode and this electrode was chosen to approximate the capacitance $C_{CG}$ of the nEDM measurement cell electrodes (see Fig. \ref{fig:Cavallo_Setup}). This created an elevated electric field region between them, necessitating additional refinement for the engineering features. This included the grooves for a poly(methyl methacrylate) (PMMA) support cylinder to support the C electrode in space, and access holes in the D electrode. Access holes were necessary to link the internal volume of the PMMA cylinder with the surrounding medium and provided locations for voltage measurement transducers. 

\subsection{General Parametric Form of the Electrode Profile Curves}

\begin{figure}
    \centering
    \includegraphics[width=0.75\linewidth]{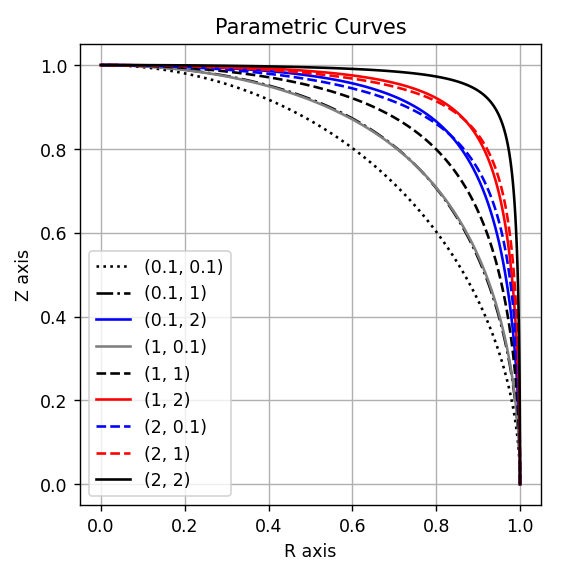}
    \caption{Examples of the electrode profile curves of Eqs. (\ref{eq:r}) and (\ref{eq:z}) from 0 to $\pi/2$ for various ($k_r,k_z$) parameters. Note how differing parameters can  result in an asymmetric distortion of the curve, favoring the elbow's displacement toward the r or $z$ axis.}
    \label{fig:Examplecurves}
\end{figure}

In this study, the basic electrode profiles were described using several smoothly joined parameterized curves defined on the positive half of a plane and then fully rotated around the vertical axis of the plane to produce axisymmetric shapes. These curves are constructed to ensure continuity and smoothness at their interfaces by converging to common limits (and their first derivatives), while still allowing curvature adjustments that help distribute the electric field lines more evenly. Since there is no perfect closed-form solution to the electrode profiles that minimizes the field over the electrode surface, we specified a generic class of parameterized equations enabling a systematic sweep. Although this parameterization may not yield a single optimal solution, it provides an efficient and flexible means of generating shapes that satisfy our design requirements.


We began with the parameterization of the equation of an ellipse:
\begin{equation}
    r = a  \cos(t)
    \label{eq:x}
\end{equation}
\begin{equation}
    z=b \sin(t),
    \label{eq:y}
\end{equation}
with $t \in [0, \pi/2]$ covering the first quadrant of the $r$-$z$ plane. The limiting slope at $t=0$ is vertical (parallel to the $z$ axis), and the slope at $t=\pi/2$ is horizontal (parallel to the $r$ axis). We join multiple curves at their $t=0$ and $\pi/2$ points to form longer profiles with continuous slope. Superellipses (Lam\`e curves) \cite{mathworld} were considered, but we found good performance with the above and the asymmetry mapping described below.

We enveloped the $r$ and $z$ formulas in Eqs. (\ref{eq:x}) and (\ref{eq:y}) in a "tanh envelope" specified below, smoothly introducing asymmetry into the geometry with the same limiting slopes (horizontal or vertical). 
Revolving these 2D closed curves around the $z$ axis creates 3D electrode shapes. The result takes the following form:
\begin{equation}
    r = R_o  \left [ \frac{\tanh(k_r\cos t)}{\tanh(k_r)} \right ] 
    \label{eq:r}
\end{equation}
\begin{equation}
    z=z_o \left [ \frac{\tanh(k_z\sin t)}{\tanh(k_z)} \right ] 
    \label{eq:z}
\end{equation}
These parametric curves contain the parameter pair ($k_r$,$k_z$), deviating from the simple ellipse that can be fine-tuned for an empirically driven electrode shape, and COMSOL features a study method to loop through sets of parameter values. 

Figure \ref{fig:Examplecurves} plots the above curves for a few values of ($k_r$,$k_z$) to demonstrate the diversity of this class of curves. The curvature is controlled by the magnitude of the parameters, and the elbow of the curve can be biased towards the $z$ or $r$ axis by choosing different ($k_r$, $k_z$) values.

\section{Results}

\begin{figure}
    \includegraphics[width=\linewidth]{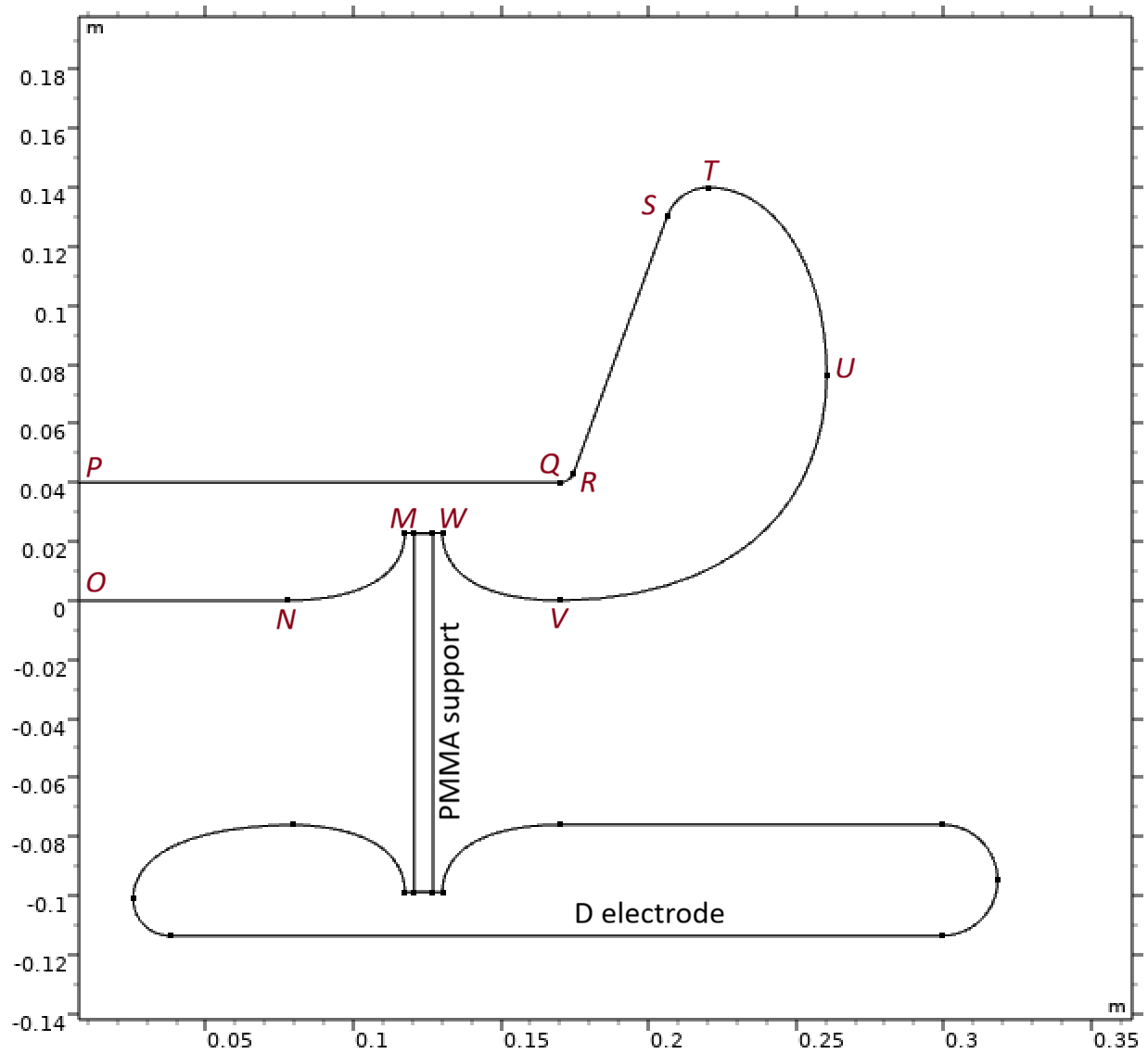}

    \vspace{0.5em}

    \fbox{
    \begin{minipage}{0.6\textwidth}
    \small
    \textbf{Legend:} \\
    \hspace*{1em} $\overline{PQ}$: Inner flat surface (horizontal line).\\
    \hspace*{1em} $\overline{QR}$: Inner bottom arc (circular 0.5 cm radius, 0\degree \hspace*{0.05em} to 20\degree).\\
    \hspace*{1em} $\overline{RS}$: 70 degree slope to join segments $\overline{QR}$ and $\overline{ST}$.\\
    \hspace*{1em} $\overline{ST}$: Inner top arc (circular 1.5 cm radius, 90\degree \hspace*{0.05em} to 160\degree). \\
    \hspace*{1em} $\overline{TU}$. Upper parametric curve. \\
    \hspace*{1em} $\overline{UV}$: Lower parametric curve. \\
    \hspace*{1em} $\overline{VW}$: Parametric curve for C electrode support groove. \\
    \hspace*{1em} $\overline{MW}$: C electrode support groove (width=1.27 cm). \\
    \hspace*{1em} $\overline{NM}$: Mirror of $\overline{VW}$. \\ 
    \hspace*{1em} $\overline{ON}$: Bottom of electrode (horizontal line). 
    \vspace{0.5em} \\
    \textbf{Important points} (where the origin is \textit{O}, the radially-centered bottom of the C electrode): \\
    \hspace*{1em} \textit{Q}: (17 cm, 4 cm) \\
    \hspace*{1em} \textit{T}: (22.177 cm, 14 cm) \\
    \hspace*{1em} \textit{U}: (26.177 cm, 7.65 cm) \\
    \hspace*{1em} \textit{V}: (17 cm, 0 cm) \\
    \hspace*{1em} \textit{N}: (7.765 cm, 0 cm)
    \end{minipage}
    }
    \normalsize
    \caption{Drawing of the C electrode cross-section within a 36.83 cm (radius) cylindrical ground return, depicted with the D electrode and support PMMA cylinder.
} \label{fig:Cshapeexplained}
\end{figure}

Our strategy for the trade-off between increasing the theoretical gain and decreasing the peak electric fields, described in detail in Sec. \ref{sec:methods} above, included shaping the A and C electrodes with tuned parametric curves, and choosing to use circular fillets for the edge of the B electrode.


Figure \ref{fig:Cshapeexplained} shows the geometry of the C electrode. For simplicity, circular arcs were chosen for segments $\overline{QR}$ and $\overline{ST}$ as labeled in the figure. Anticipating a high electric field on the outside of the C electrode, we formed the lobe with a balance of two parametric curves, $\overline{TU}$ and $\overline{UV}$. The description of these two segments follow Eqs. (\ref{eq:r}) and (\ref{eq:z}), with $t \in[0,\pi/2]$. The coefficients $R_o$ and $z_o$ for each curve can be viewed as extending the equation over the ranges from points \textit{T} to \textit{U}, and from points \textit{U} to \textit{V}, respectively. Given the geometric constraints of the volume containing the electrodes -- a ground return radius of 36.83 cm and a vertical height of about 50 cm -- an earlier iteration set the vertical height of the lobe, $z_{B}$, to 14 cm, and the horizontal component of $\overline{TU}$, to 4 cm. In addition to the free parameters in Eqs. (\ref{eq:r}) and (\ref{eq:z}), adjusting the $z$ coordinate where the two curves meet, $z_C$, introduced an extra degree of freedom to better distribute the electric field lines. The coordinates of selected points are listed in the caption of Fig. \ref{fig:Cshapeexplained}. The coefficients and the $k$ parameters ($k_{r}$, $k_{z}$) for both curves are summarized in Table \ref{tab:params} under $C_{\overline{TU}}$ and $C_{\overline{TU}}$ . 

\begin{figure}
    \centering
    \includegraphics[width=0.75\linewidth]{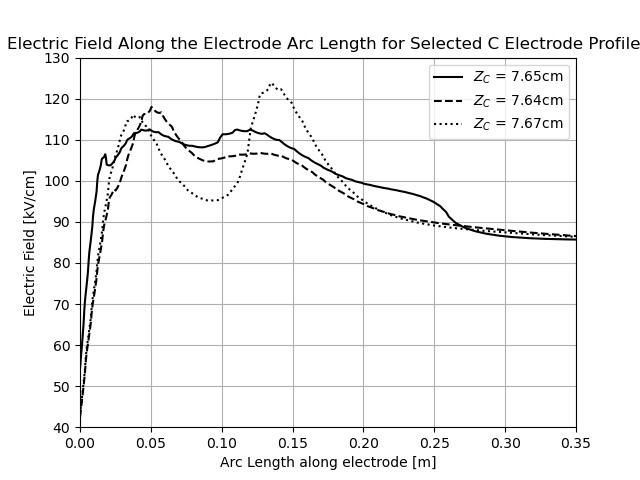}
    \caption{Electric Field Profile as a function of arc length along the fully-charged C electrode (650 kV). The arc length is calculated by tracing the surface of the electrode from points \textit{S} to \textit{V} in Fig \ref{fig:Cshapeexplained}. The solid line, $z_C$ = 7.65 cm, corresponds to the chosen parameters for our design.}
    \label{fig:CArcLength}
\end{figure}

Figure \ref{fig:CArcLength} illustrates the results of varying $z_C$ as part of our parameter sweep to reduce electric field maxima along the outer lobe of the C electrode. This figure shows the electric field strength on the surface of the C electrode as a function of arc length, obtained by following the surface of the electrode from point S through point V in Fig \ref{fig:Cshapeexplained}. The solid line corresponds to our chosen profile, contrasted with two other parameter sets with slightly different values of $z_C$ (dotted lines) with higher electric field peaks. Tuning the shape of the electrodes changes the breakdown probabilities (see Sec. \ref{sectionA} for breakdown probability discussion) in LHe at 12 Torr from 92.5\%  for $z_{C}$=7.64 cm or 94.1\% for 7.67 cm, to 92.1\% for $z_{C}$=7.65. Although the lightest dotted curve ($z_{C}$=7.64 cm) shows the lowest discharge probability, the solid curve was selected for its comparable probability and lower electric field maximum.

\begin{table}
    \centering
    \begin{tabular}{ | c | cccc |}
        \hline Name & $R_o$ & $z_o$ & $k_r$ & $k_z$ \\ \hline
        $C_{\overline{TU}}$ & 4.000 cm & 6.35 cm & 0.2 & 0.3 \\
        $C_{\overline{UV}}$ & 9.177 cm & -7.65 cm & 0.5 & 0.4 \\
        $A_{inner}$ & -4.320 cm & -1.25 cm & 0.8 & 3.0 \\
        $A_{outer}$ & -4.320 cm & -1.25 cm & 1.0 & 1.0 \\
        C-D Groove & 3.983 cm & 2.283 cm & 0.5 & 0.5 \\
        D holes & 5.643 cm & 2.508 cm & 0.5 & 0.5 \\ \hline
    \end{tabular}
    \caption{Parameters for the electrode curves for a Cavallo multiplier design within a 36.83 cm (radius) cylindrical ground return. For k values less than 1, the resulting shapes are essentially simple ellipses, as the asymmetry function has very little effect.}
    \label{tab:params}
\end{table}

\begin{figure}
    \includegraphics[width=\linewidth]{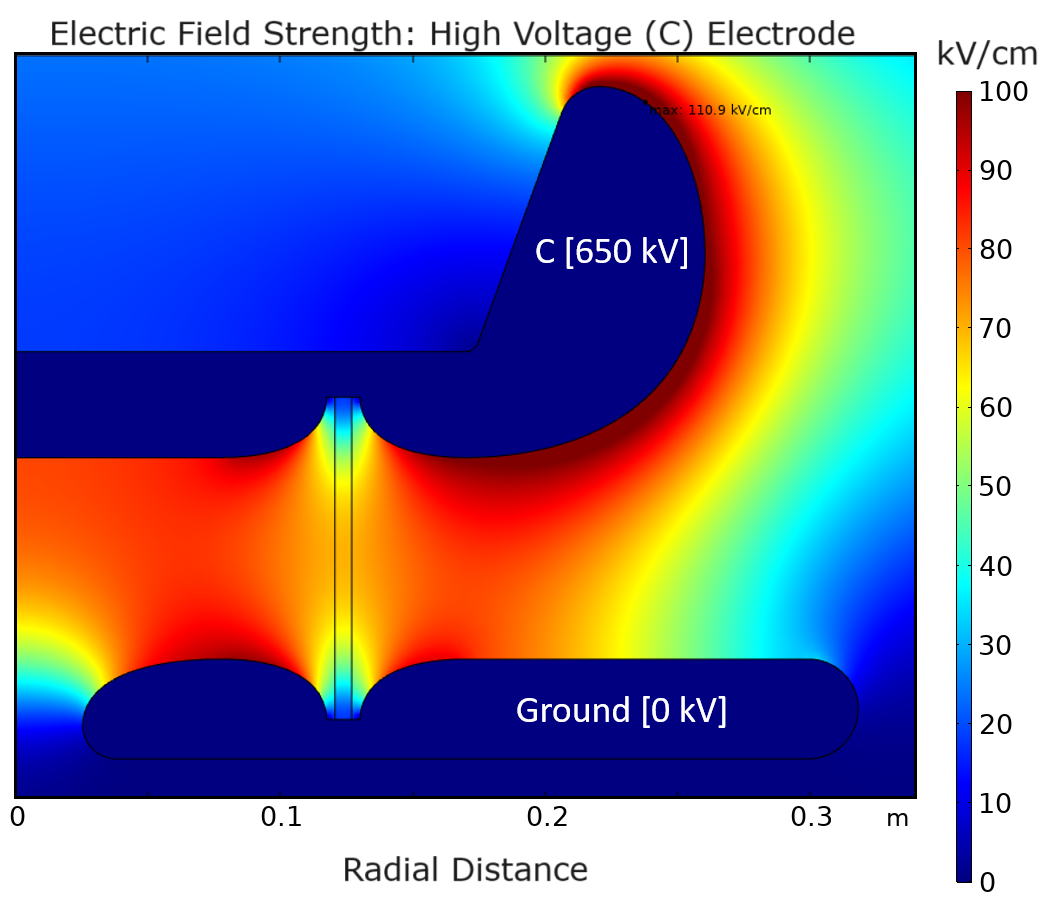}
    \caption{Electric field map of the C electrode in the test stand, held up by a PMMA cylinder from a ground electrode (electrode D). The engineering surfaces, including a hole in the D electrode (Ground), all employ the parametric curves. } \label{fig:CDeng}
\end{figure}

A hollow PMMA cylinder supports the C electrode, separating it from the D electrode at a distance of 7.6 cm to approximate the "load" of a measurement cell capacitance. The edges of a dielectric must hide in a low field region \cite{ito_apparatus_2016}; the PMMA cylinder mates with both electrodes in a groove carved out by a curve mirrored about the center of the groove (segment $\overline{MW}$ in Fig. \ref{fig:Cshapeexplained}). It shares the form of Eqs. (\ref{eq:r}) and (\ref{eq:z}), but parameterized from $t\in[-\pi/2,0]$ and maximizes the curvature space available, with $R_o$ as the distance from the inside of the groove to the beginning of the outer lobe (3.98 cm), and $z_o$ as the depth of the groove (2.28 cm deep) needed to shield the PMMA's edges. The tuned parameters for this are labelled "C-D Groove" in Table \ref{tab:params}. 

Furthermore, a parametric curve was also employed for access holes in the D electrode, to make measurements of the C electrode voltage using a field mill as done in Ref. \cite{maruvada_development_1983}. Here the coefficients were parameterized as a tunable fraction of the thickness of the D electrode, $h_D$=3.76 cm, and are listed in Table \ref{tab:params} under "D holes."


\begin{figure}
  \begin{subfigure}{0.5\textwidth}
  \centering
    \includegraphics[width=\linewidth]{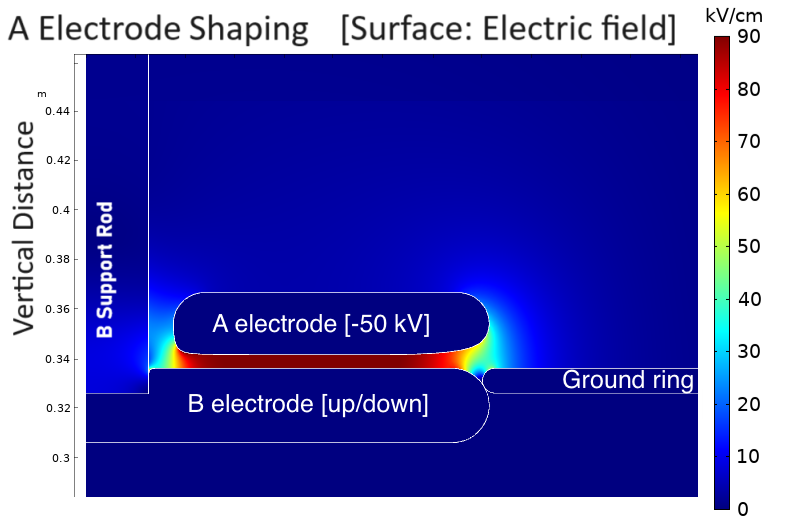}
    \caption{} 
  \end{subfigure}%
  \hspace*{\fill}   
  \begin{subfigure}{0.5\textwidth}
  \centering
    \includegraphics[width=\linewidth]{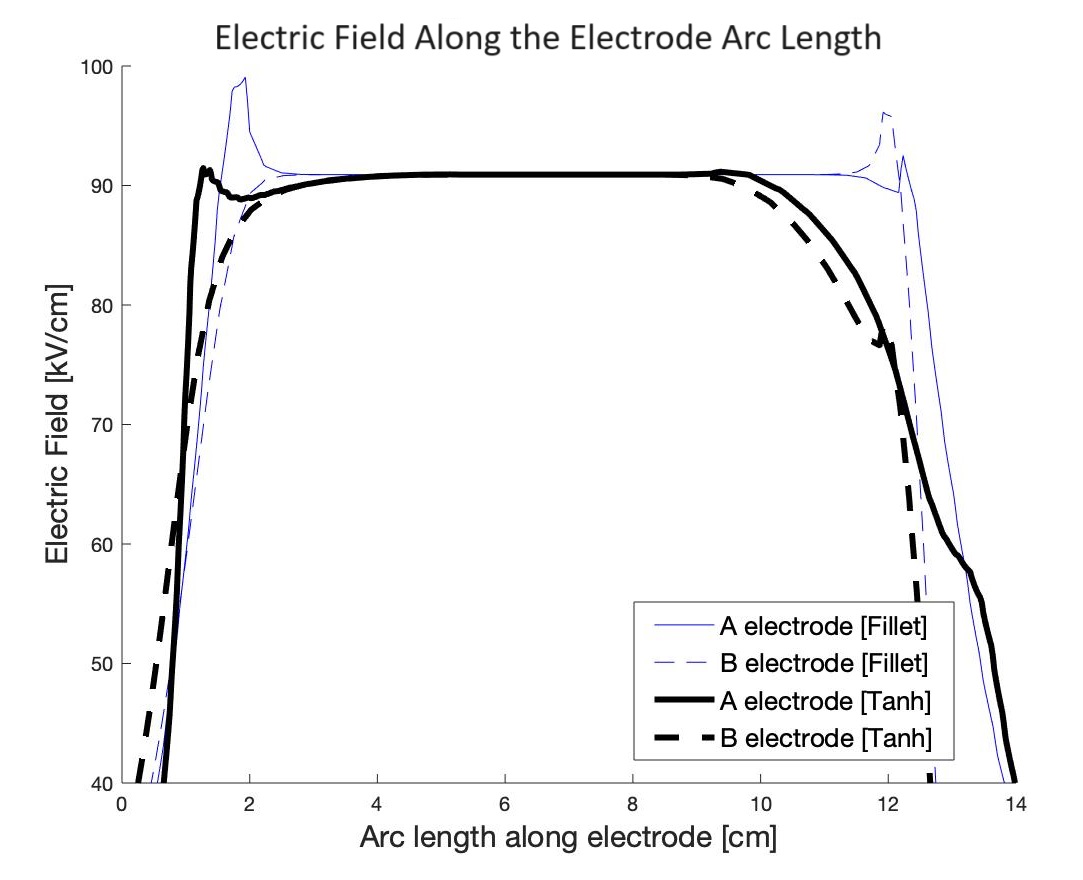}
    \caption{} 
  \end{subfigure}%
\caption{The strong electric field between the A and B induces a large charge on the B electrode, but any electric field enhancements may result in electrical discharge. Left (a): Axisymmetric simulation of the electric field between the A and B electrode. Parametric curves decreased the maxima on the electrode edges.
Right (b): Surface electric field of the A and B electrodes plotted as a function of arc length along the electrode's cross-sectional edge. The A electrode's arc length is measured from the innermost point of the A electrode (the leftmost in Figure a), and the B arc length is measured from the bottom corner of the B electrode shaft's borehole. The light colored lines representing the fillet edge treatment produce higher fields than those terminated with the parametric curves, denoted "Tanh" for the hyperbolic tangent functions.} 
\label{fig:AB}
\end{figure}

The bottom surface of the A electrode experiences a high electric field when the B electrode is docked on the ground ring as shown in Fig. \ref{fig:AB}(a). Separate parameterized curves are needed for the inner and outer curves of the A electrode because the ground ring and B electrode shaft distort the electric field. Both take the form of Eqs. (\ref{eq:r}) and (\ref{eq:z}) from $t \in[0,\pi/2]$, where the coefficients ($R_o,z_o$) were parameterized with a fraction of the radius and height of the A electrode, respectively. A percentage of 0.25 was sufficient for both inner and outer curves. The resultant coefficients and optimal ($k_r,k_z$) pairs are listed in Table \ref{tab:params} labeled "$A_{inner}$" and "$A_{outer}$" for the inner and outer curves, respectively. 

Comparing these curves with circular fillets highlights their effectiveness against the electric field enhancements between A and B. Figure \ref{fig:AB}(b) is a line graph showing the electric field strength along the surface of the stressed edges of the A and B electrode cross sections from Fig. \ref{fig:AB}(a), plotted by position along the arc length of each electrode. As expected, the circular fillet (graphed in blue) results in two large electric field maxima/peaks at arc length 2 cm and 12 cm. Conversely, the tailored parametric curves (denoted "Tanh" in the legend) eliminated the peak on the outer edge and reduced the inner peak to less than 91.5 kV/cm (moved to 1.3 cm due to the arc shape). 

\begin{figure}
    \includegraphics[width=\linewidth]{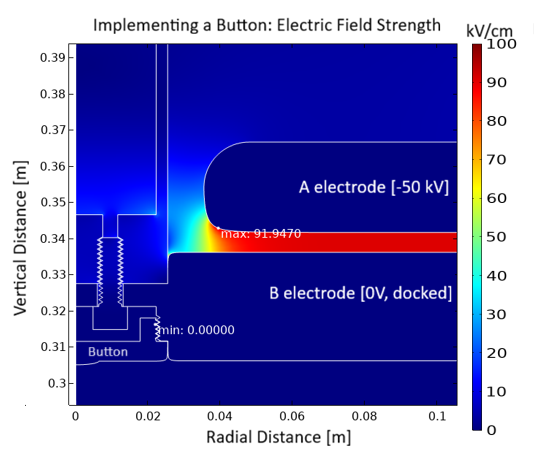}
    \caption{Details of the B electrode: Here we can see that a sparking button can also hide the hardware surfaces such as shaft screws for the B electrode. Note, however, that these screws are suggestive only, and are not correct in the 2D-axisymmetric framework.} \label{fig:button0}
\end{figure}

Despite their demonstrated utility, the parametric curves were not used on all features. Parametric curves for some of these surfaces (e.g., sides of the button and its recessed pocket in the B electrode in Fig. \ref{fig:button0}, or the interface of the B electrode with its driving rod) were considered, but these surfaces did not make sufficiently large enough contributions to the breakdown probability to warrant the expense and fabrication effort. 

\begin{figure}
    \includegraphics[width=\linewidth]{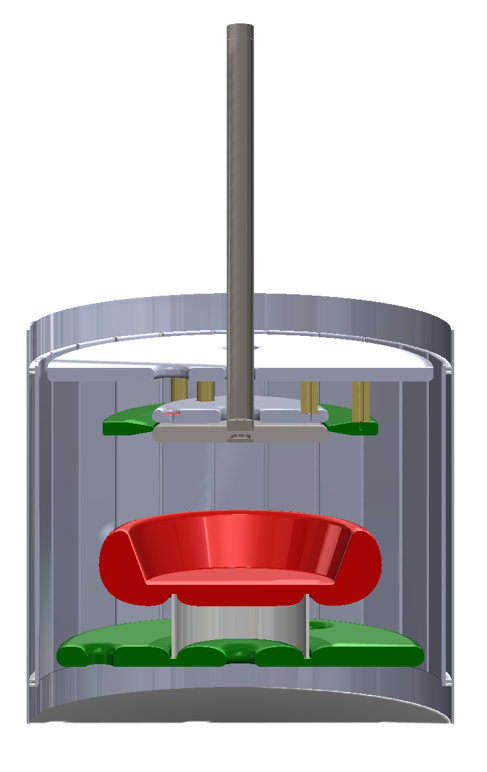}
    \caption{Final Computer-Aided Design: The electrodes are shown with their engineering surfaces, as well as a ground return made of 24 slats.} \label{fig:Solidworks}
\end{figure}

\begin{figure}
  \begin{subfigure}{0.5\textwidth}
    \includegraphics[width=\linewidth]{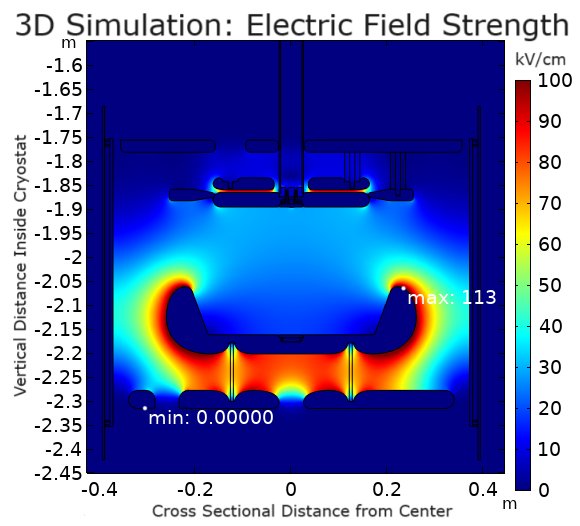}
    \caption{Electric Field Cross Section for 3D Apparatus}
  \end{subfigure}%
  \hspace*{\fill}   
  \begin{subfigure}{0.5\textwidth}
    \includegraphics[width=\linewidth]{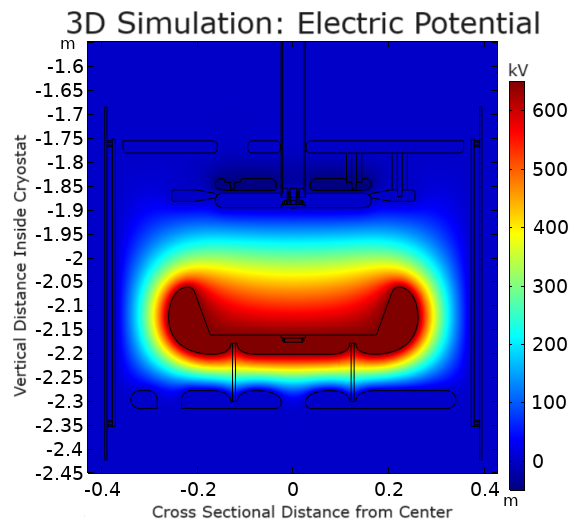}
    \caption{Voltage Cross Section for 3D Apparatus} 
  \end{subfigure}%
\caption{3D cross-section simulation: Figure \ref{fig:Solidworks} was imported into the finite element analysis program, and the running voltages of $V_A=-50 kV$ and $V_C=650$ kV were applied} \label{fig:3D}
\end{figure}

Our final design is depicted in Fig. \ref{fig:Solidworks}, with the high voltage electrode shown in red. The ground return on the outer wall is created using 24 flat slats to reduce eddy current heating and facilitate access to the inner volume. The COMSOL simulation in Fig. \ref{fig:3D} shows that the total engineering surfaces work as designed. The final mathematical gain is 18, and our final maximum electric field is 116 kV/cm. The apparatus will reach our 650 kV target at around 14 cycles. 

\section{Analysis}
\label{sectionA}

\begin{figure}
    \includegraphics[width=\linewidth]{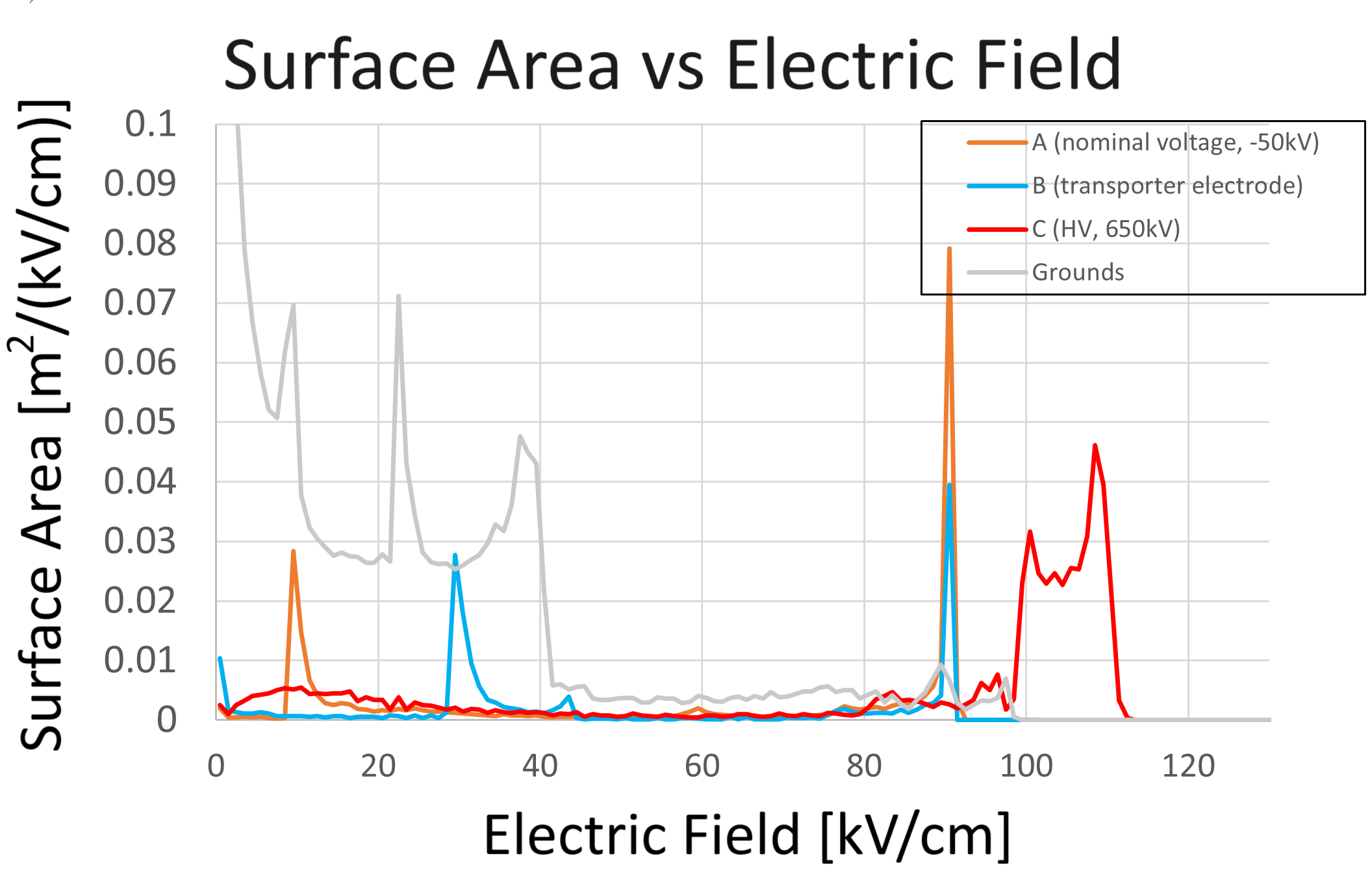}
    \caption{Electrode surface area as a function of electric field strength. The finite element analysis program integrated through each surface area element in Fig. \ref{fig:3D}(a), and histogrammed that surface area by its surface electric field strength.} \label{fig:SA}
\end{figure}

Electrical breakdown is often conceptualized as a simple electric-field threshold effect, and it is straightforward to constrain the maximum electric field in our design to less than 120 kV/cm. However, a comprehensive analysis of electric field breakdown in LHe done by Phan et al. \cite{phan_study_2021} showed that the breakdown probability is not governed by a single threshold. Instead, the probability depends on the integrated electrode surface area exposed to a given electric field strength:
\begin{equation}
    P_{breakdown} = 1 - \prod_{E_i} e^{-S(E_i) W(E_i)}
    \label{eq:Nguyen}
\end{equation}
where $S(E_i)$ is the surface area experiencing the electric field strength $E_i$. Factors affecting the breakdown probability such as the electrode roughness or the pressure of the medium are grouped  into the hazard function, $W(E_i)$.
By using a previously measured $W$ curve -- data taken from a set of simple electrodes with similar electrode material and polishing, and at the same helium pressure -- we can apply Eq. (\ref{eq:Nguyen}) to calculate a breakdown probability for reaching a specific voltage for an arbitrary electrode geometry. Figure \ref{fig:SA} shows the electric field distribution for our specific electrode geometry at the fully charged configuration in Fig. \ref{fig:3D}(a); the electrode surface area is characterized by the electric field strength at the surface, and then discretized into the nearest kV/cm bin. 

The polarity of the electric field needs to be reversed periodically for an nEDM experiment. The electric field distribution on the high voltage electrode differs significantly from that of the ground electrodes. Therefore, understanding whether breakdown initiates on a particular polarity or on either electrode indiscriminately is important to correctly calculating the probability using eq. (\ref{eq:Nguyen}).

Previous studies of the polarity effect in LHe show that breakdown strength depends strongly on cathode material and surface condition, particularly the metal work function, implying that breakdown is governed by the electric field at the cathode surface rather than at the anode \cite{Fujii_1979,YOSHINO1982305,YOSHINO4156770}. The similarity between the hazard function $W$ and the Fowler–Nordheim field-emission function further suggests that breakdown is always initiated by field emission from the cathode \cite{phan_study_2021}. Under this assumption, when the C electrode is negatively charged, only the surface area of the C electrode is relevant. Even if breakdown initiation at the anode is allowed, the contribution from the C electrode still dominates, since most of the anode surface area (grounds) lies in the low-field region, whereas a large fraction of the C electrode surface experiences fields between 100 and 113 kV/cm. Including the total surface area of all electrodes therefore provides a conservative upper bound on the breakdown calculation. Furthermore, the geometric asymmetry between the ground and C electrode can make the breakdown initiation surface apparent. If the discharge must originate at the cathode, then charging the C electrode to a positive voltage results in a significantly lower breakdown probability. In contrast, if breakdown can originate at either electrode, the predicted breakdown probability is identical for positive and negative charging configurations.

Following the process by Phan et al.\cite{phan_study_2021} with the data obtained for stainless steel electrodes, we calculated the predicted breakdown probabilities for the fully charged Cavallo apparatus using the surface area field distributions shown in Fig. \ref{fig:SA}. The resulting probabilities are summarized in Table \ref{tab:breakdowns}. They indicate that, while the probability is sufficiently low at 1520 Torr, the apparatus is likely to break down at low pressure with high probability. Experimental studies at 600 torr and 1520 torr, initially employing electropolished stainless steel electrodes, will not only verify these calculations, but also inform charging practices and strategies for minimizing damaging breakdowns during charge transfer at the B-A ground-ring contact and the B–C contact button.

\begin{table}
    \centering
    \begin{tabular}{ | c | c  c | }
        \hline
        & C electrode & All Surfaces \\ 
        \hline
         LHe at 12 torr * & 98.2\% & 99.3\% \\  	
         LHe at 600 torr & 0.717\% & 0.783\% \\ 
         LHe at 1520 torr & 0.0003\% & 0.0003\% \\ 
         \hline
    \end{tabular} \\
        * This corresponds to the saturated vapor pressure of LHe at $\approx$1.7K.
    \caption{Predicted probabilities for the Cavallo apparatus breakdown}
    \label{tab:breakdowns}
\end{table}


\subsection{The B-C Charge Transfer Breakdown \label{section:BCbreakdown}}


Charge transfer from the B to C electrodes may involve electrical breakdown confined to their reinforced, replaceable buttons. To limit damage, the breakdown energy must remain low; these buttons were designed to undergo breakdown only at a reasonably small electrode separation. This was verified by correlating the breakdown energy with the probability of breakdown.

\begin{figure}
  \begin{subfigure}{0.5\textwidth}
    \includegraphics[width=\linewidth]{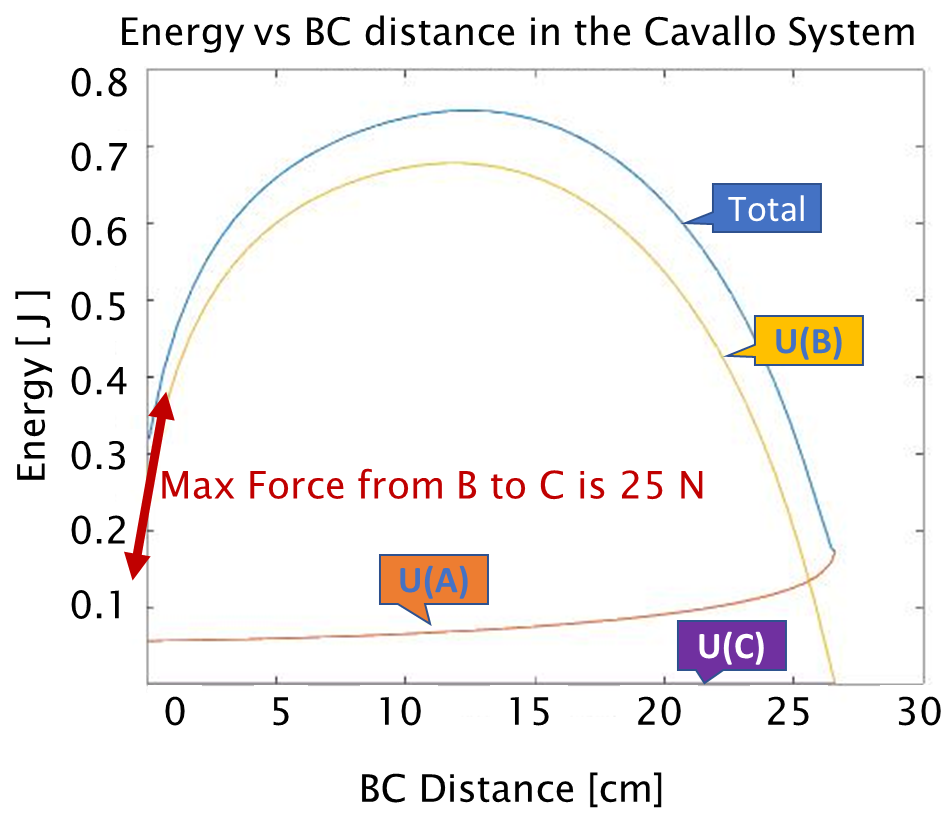}
    \caption{} 
  \end{subfigure}%
  \hspace*{\fill}   
  \begin{subfigure}{0.5\textwidth}
    \includegraphics[width=\linewidth]{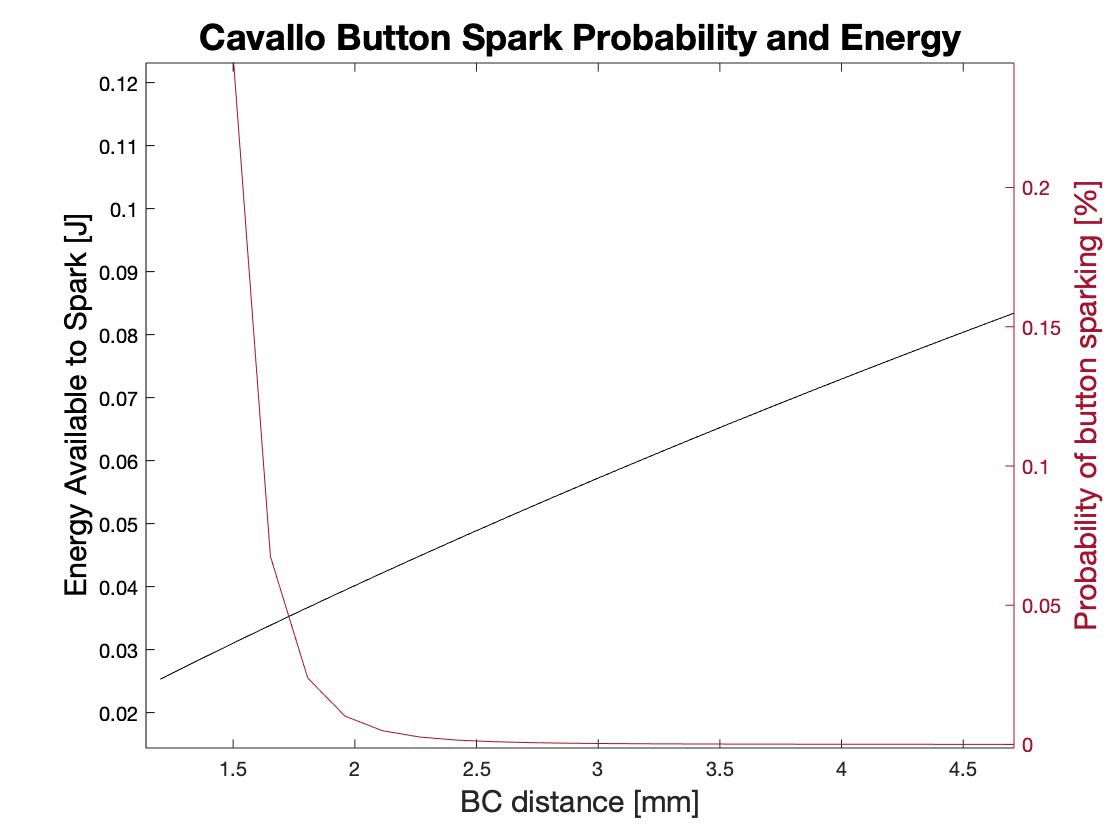}
    \caption{} 
  \end{subfigure}%
\caption{Potential energies in the first charge cycle of the Cavallo apparatus. Left (a): Total electrostatic energy of each electrode as a function of the distance between the B and C electrodes. The force between the B and C electrodes is the derivative of U(B)-U(C), calculated to be less than 25 N. Right (b): Electrical discharge profile. The black curve (left axis) is the amount of energy available to a spark between the B and C electrodes, as a function of the distance between the B and C electrodes. The red curve (right axis) is the breakdown probability between the B and C electrodes at the fixed distance. The closer the electrodes become, the lower amount of energy available to the spark; the probability curve becomes non-negligible when the B-C electrode distance is less than 2 mm, with only 35 mJ available.} \label{fig:Energies}
\end{figure}

In Fig. \ref{fig:Energies}(a), the total energy of the system and the energy of each electrode was calculated using $U = \frac{1}{2} QV$ and plotted as a function of the B electrode position as it decreases the distance to the C electrode. However, only the change in potential energy before and after charge transfer is available for the breakdown, assuming the process fully eliminates the potential difference between the B and C electrodes.
The result is the black curve in Fig. \ref{fig:Energies}(b). Using the electric field between electrodes B and C, an estimate of the breakdown probability exponentially increases for B-C distances below 2mm, corresponding to an energy of 0.035 J. 

This breakdown probability curve was calculated as follows. The surface area distribution vs. electric field strength $S(E_i)$ was computed for incremental steps of the B electrode toward C, using the initially loaded charge on B. This initial charge corresponds to the first stroke of the Cavallo apparatus, for which the charge difference between B and C (and hence the available spark energy) is maximal. Then, Eq. (\ref{eq:Nguyen}) was applied. However, this equation was originally derived for a voltage ramp supplied by an external power source rather than for an in situ voltage change arising from geometric evolution. Nevertheless, we expect the application of this equation and its associated hazard function $W$ to remain applicable, as the C electrode voltage (which dominates the breakdown probability) increases monotonically during the approach. Further experimental tests of the contact-button design would be valuable to validate this assumption.


\begin{figure}
    \includegraphics[width=\linewidth]{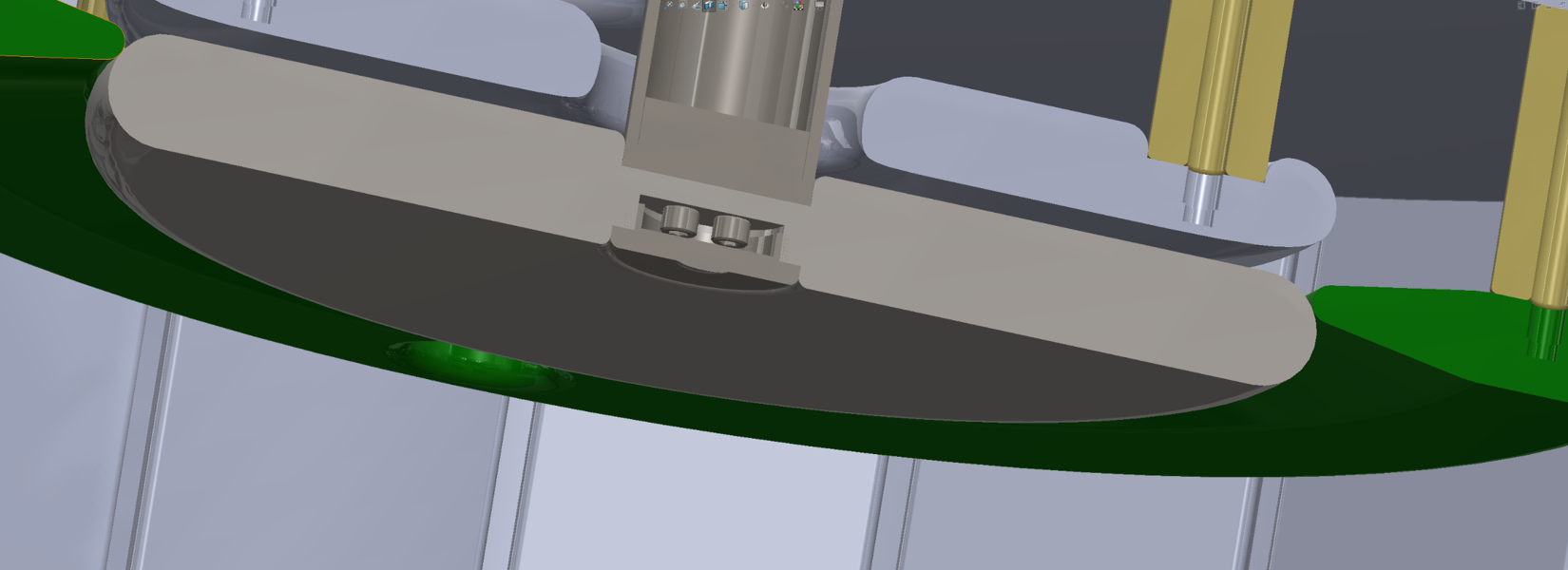}
    \caption{A Computer-Aided Drawing of the spark button and the shaft hardware for the B electrode.} \label{fig:CADbutton}
\end{figure}

\begin{figure}
    \includegraphics[width=\linewidth]{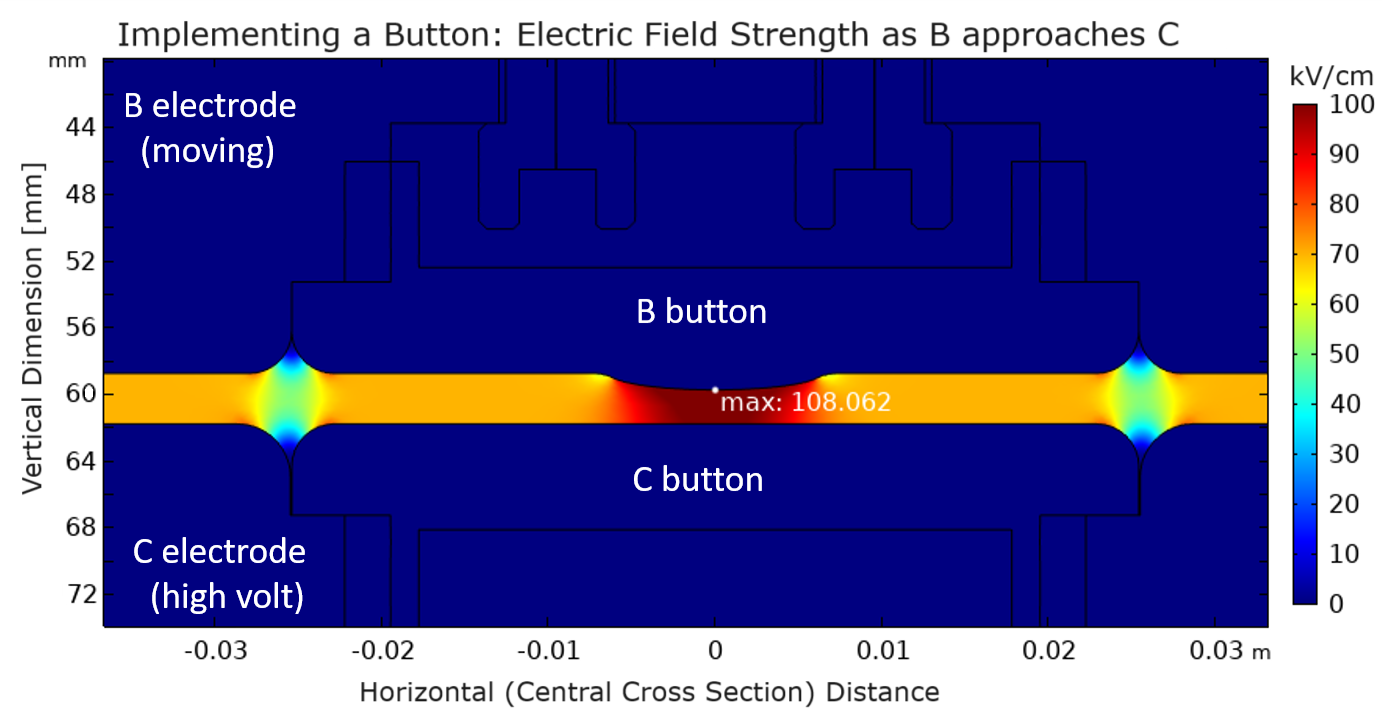}
    \caption{Sparking buttons installed on both the B and C electrodes---thickened and replaceable---protect the electrodes as a sacrificial region. Note the high electric field to force the spark in the desired location.} \label{fig:button1}
\end{figure}

Electric field simulations were performed for the configuration corresponding to a 2 mm electrode separation, below which breakdown probability increases nonlinearly, with the results shown in Fig. \ref{fig:button1}. The hightest electric field occurs at the button, as designed for the discharge.

An added benefit to the button is that it can hide necessary engineering features. For example, shown in Fig. \ref{fig:CADbutton} on the B electrode, the button electrostatically hides the hardware connecting B to an actuator-driven rod. 

\subsection{Applicability to the nEDM experiment configuration}

\begin{figure}
    \includegraphics[width=\linewidth]{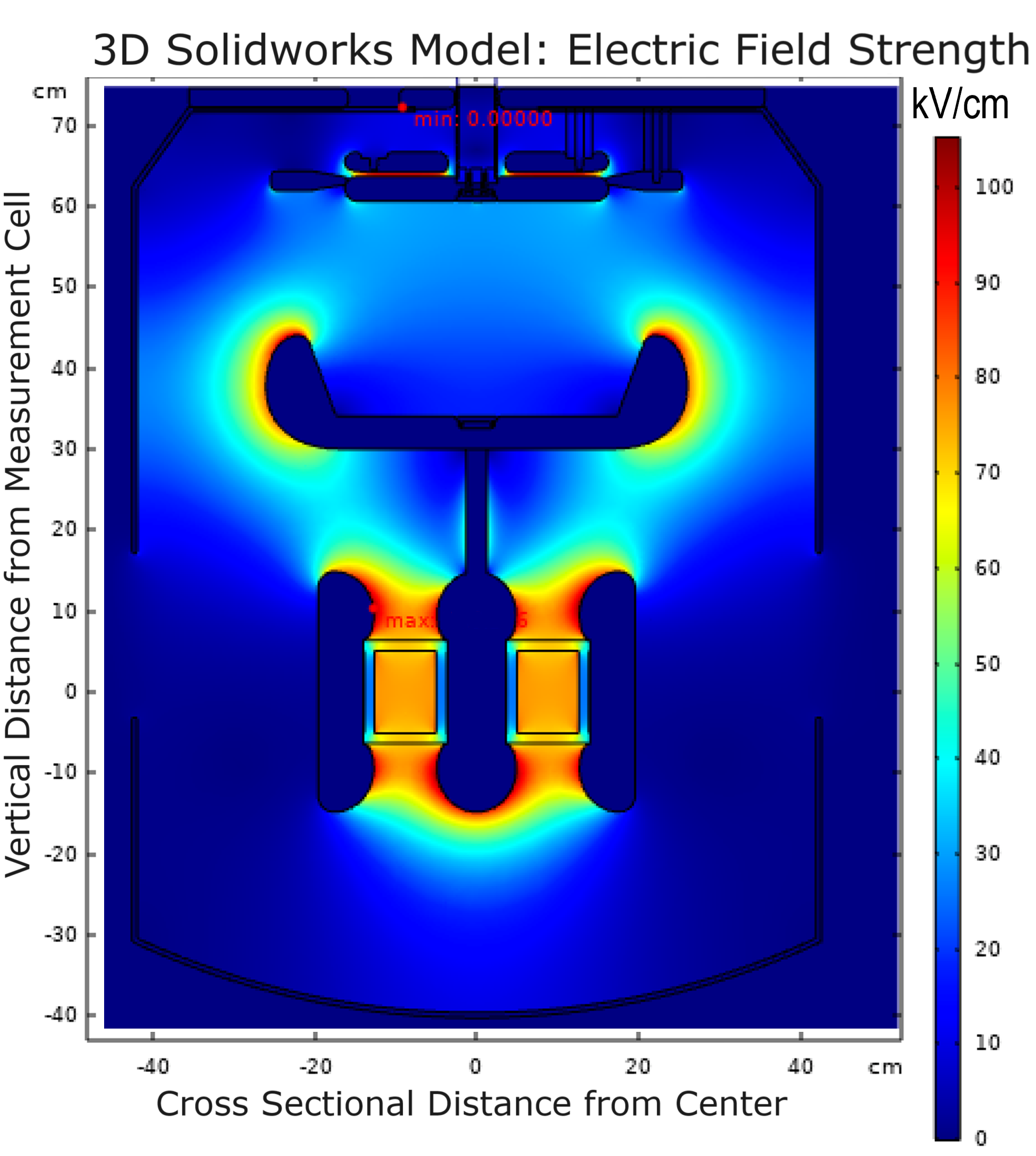}
    \caption{Electric Field in the Future nEDM Experimental Volume: 3D Model cross-section of the Cavallo Electrodes in their space in the final nEDM experiment drawn on the right in Fig. \ref{fig:Cavallo_Setup}, with the measurement electrodes attached below.} \label{fig:nEDM}
\end{figure}

Following the test apparatus simulation results, these electrodes were then modeled in their position in the nEDM experimental apparatus, shown in Fig. \ref{fig:nEDM}. Note that the nEDM experimental apparatus gives the Cavallo electrodes more space, resulting in lower electric fields than those in Fig. \ref{fig:3D}(a). Per the design requirements, the electric field is on the order of the test apparatus ({~}{{\raisebox{-.25em}{\textasciitilde}}}{120} kV/cm) around the measurement cells.

\begin{figure}
    \includegraphics[width=\linewidth]{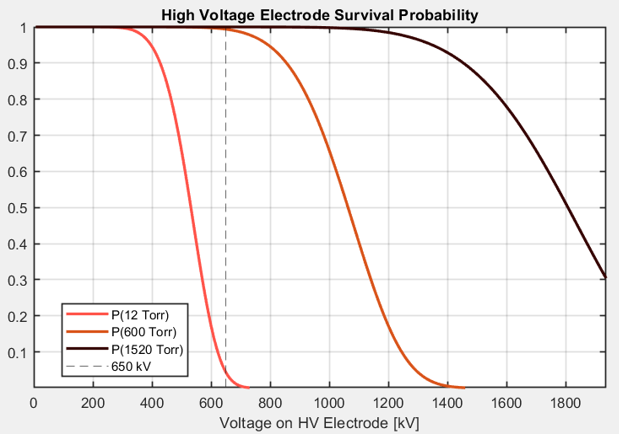}
    \caption{High Voltage Electrode Survival Probability: As the ramped voltage increases on the high voltage electrodes (the measurement and C electrode connected together as in Fig. \ref{fig:nEDM}), the probability of reaching that voltage without electrical breakdown is calculated (assuming electropolished stainless steel electrodes), to create these probability curves. } \label{fig:Survive}
\end{figure}

Figure \ref{fig:Survive} shows the survival probability of the nEDM high voltage electrodes as a function of the C electrode voltage. This probability is defined as the probability of the electrode design to reach the specified C electrode voltage without breakdown, calculated as $1- P_{breakdown}$ [where $P_{breakdown}$ is defined by Eq. (\ref{eq:Nguyen})].  The calculation used the hazard ($W$) function for stainless steel for the electric field configuration shown in Fig. \ref{fig:nEDM} (which includes both the Cavallo high voltage and the nEDM cell high voltage electrodes as depicted in Fig \ref{fig:Cavallo_Setup}). While stainless steel cannot be used in the final experiment, the suitable materials discussed in the introduction are likely to have a similar $W$ function to the stainless steel electrodes \cite{itoHV_now}; this will be verified by measurement. 
In this calculation, the electrode design has a high survival probability at voltages well above the target 650 kV at the nEDM experimental design pressure of 1520 Torr. The probability of breakdown [Eq. (\ref{eq:Nguyen})] for the system to ramp to 650 kV is on the order of $10^{-6}$; Therefore, we should be able to achieve this design voltage with the electrode shapes obtained through the methods discussed in Sec.~\ref{sec:methods}.

\section{Conclusions}

The electrode shapes for the Cavallo apparatus were fine-tuned using parametric hyperbolic-tangent-based curves within finite element analysis simulations. This allowed us to spread out the electric field lines and mitigate regions of electric field peaks. The results of this design meet the goals: a geometry-defined gain of 18 allows us to reach the high voltage goal of 650 kV in 14 cycles. When the B electrode delivers the charge to the C electrode, any resulting spark in the charge transfer is limited to the order of 10 mJ on a reinforced replaceable button.

These shapes were incorporated into the engineering of both a test apparatus and the final nEDM experimental design. The engineering design was then modeled in the finite element analysis program, and analyzed for a probability of breakdown. These calculations show that the Cavallo electrodes have a suitably low breakdown probability in pressurized LHe.


\section{Acknowledgments}

Thanks to Stefan Baessler and Chris Crawford, whose correspondence was invaluable to this work. This work was supported by the United States Department of Energy, Office of Science, Office of Nuclear Physics under contract number 89233218CNA000001 under field work proposal LANLEEDM, the Los Alamos National Laboratory Integrated Contract Order No. 4000129433 with Oak Ridge National Laboratory, and the National Science Foundation Grants NSF-2110898, NSF-1822515, and NSF-1812340---“Fundamental Studies in Nuclear Physics.”

\bibliographystyle{elsarticle-num-names}
\bibliography{CavalloBibtex.bib}






\end{document}